\documentclass[final]{aa}

\usepackage{txfonts}
\usepackage{times}
\usepackage{latexsym}
\usepackage{amssymb}
\usepackage{flafter}

\usepackage{natbib}
\bibpunct{(}{)}{;}{}{}{,}

\usepackage{calc}
\usepackage{graphicx}
\usepackage[english]{babel}

\newcommand{\eg}{e.g.{}}
\newcommand{\ie}{i.e.{}}
\newcommand{\degree}{\ensuremath{^\circ}}

\renewcommand{\arcsec}{\ensuremath{^{\prime\prime}}}

\newcommand{\hmspt}[4]{\mbox{#1\ensuremath{^\mathrm{h}}#2\ensuremath{^\mathrm{m}}#3\mbox{\rlap{.}$\mathrm{^s}$}#4}}
\newcommand{\dms}[4][]{\mbox{$#1$#2\ensuremath{^\circ}#3\ensuremath{^\prime}#4\ensuremath{^{\prime\prime}}}}
\newcommand{\eqeighttooo}[8]{\mbox{$\alpha(2000)=\hmspt{#1}{#2}{#3}{#4}$}, \mbox{$\delta=\dms[#5]{#6}{#7}{#8}$}}

\newcommand{\Msun}{\ensuremath{\mathrm{M}_\odot}}

\newcommand{\pmag}{\ensuremath{\mathrm{mag^{-1}}}}

\newcommand{\pcms}{\ensuremath{\mathrm{cm}^{-2}}}

\newcommand{\pcmc}{\ensuremath{\mathrm{cm}^{-3}}}

\newcommand{\kmps}{\ensuremath{\mathrm{km\,s^{-1}}}}
\newcommand{\K}{\ensuremath{\mathrm{K}}}

\newcommand{\pc}{\ensuremath{\mathrm{pc}}}
\newcommand{\chisquare}{\ensuremath{\chi^2}}
\newcommand{\chisq}{\chisquare}

\newcommand{\DVtherm}{\ensuremath{\Delta V_\mathrm{therm}}}

\newcommand{\DV}{\ensuremath{\Delta V}}

\newcommand{\Htwo}{\mbox{H$_2$}}

\newcommand{\NHt}{\mbox{NH$_3$}}
\newcommand{\CeiO}{\mbox{C$\ensuremath{\:\!}^{18}\ensuremath{\!\;\!}$O}}
\newcommand{\thCO}{\mbox{$\ensuremath{\:\!}^{13}\ensuremath{\!\;\!}$CO}}

\newcommand{\chin}{\ensuremath{\chi(\NHt)}}

\newcommand{\NNHt}{\ensuremath{N(\NHt)}}

\newcommand{\EHK}{\ensuremath{E(H-K)}}

\newcommand{\Fnu}{\ensuremath{F_\nu}}

\newcommand{\Tot}{\ensuremath{T_{12}}}
\newcommand{\Tex}{\ensuremath{T_\mathrm{ex}}}
\newcommand{\Tb}{\ensuremath{T_\mathrm{b}}}
\newcommand{\nc}{\ensuremath{n_\mathrm{c}}}
\newcommand{\Nc}{\ensuremath{N_\mathrm{c}}}

\newcommand{\PR}{\ensuremath{P_\mathrm{R}}}

\newcommand{\Vlsr}{\ensuremath{V_\mathrm{lsr}}}

\newcommand{\oneone}{\mbox{$(J,K)=(1,1)$}}
\newcommand{\twotwo}{\mbox{$(J,K)=(2,2)$}}
\newcommand{\onetwo}{\mbox{$(J,K)=(1,1)$} and \mbox{$(2,2)$}}
\newcommand{\onezero}{\mbox{$(J=1\mbox{--}0)$}}

\newcommand{\rv}[1]{\textbf{\small#1}}
\renewcommand{\rv}[1]{#1}
\newcommand{\expm}[2]{\ensuremath{{#1}\times10^{#2}}}

\begin{document}
   \title{The kinetic temperature of Barnard~68 \thanks{Based on observations 
          with the 100-m telescope of the MPIfR (Max-Planck-Institut 
          f\"ur Radioastronomie) at Effelsberg.}}

   \author{S.~Hotzel
          \and
          J.~Harju
          \and
          M.~Juvela
          }

   \offprints{S.~Hotzel, \email{hotzel@astro.helsinki.fi}}
   \institute{Observatory, P.O.~Box~14, FIN-00014 University of Helsinki, Finland}
   \date{Received 12 September 2002 / Accepted 26 September 2002}

   \abstract{We have observed
   the nearby isolated globule Barnard~68 (B68) in the \onetwo\
   inversion lines of ammonia. The gas kinetic temperature derived
   from these is 
   \mbox{$T=10\pm1.2~\K$}. The observed line-widths are almost
   thermal: \mbox{$\DV=0.181\pm0.003~\kmps$}
   (\mbox{$\DVtherm=
   0.164\pm0.010~\kmps$}),  
   supporting the earlier hypothesis that B68 is in hydrostatic
   equilibrium. 
   The kinetic temperature is an input
   parameter to the physical cloud model put forward recently, and we
   discuss the impact of the new value in this context.
   }

   \maketitle

\section{Introduction\label{sec:introduction}}

Discovered by \citet{barnard19}, the isolated, starless
globule Barnard~68 (B68)
received increased attention recently, after \citet{alves_01}
presented a high resolution extinction map and a cloud
model, suggesting that B68 has the physical structure of a so-called
Bonnor-Ebert sphere (BES, \ie\ a pressure bound, isothermal sphere in
hydrostatic equilibrium).
With its column density and thus number density profiles
revealed, B68 became an ideal object to study 
molecular depletion in dark
core interiors \citep[][hereafter Paper~I]{bergin_02,hotzel_02},
molecular abundances 
for a number of species \citep{difrancesco_02} and 
the gas-to-dust ratio (Paper~I).

The BES cloud model as based on the measurements of \citet{alves_01}
fixes the normalised profiles of the 
density ($n/\nc$) and the column density ($N/\Nc$) as functions of the
normalised radius ($r/R$), while for the absolute values of 
the central density \nc, the central column density \Nc\ and the radius
$R$
additional measurements are necessary.
The knowledge on any two of the parameters \nc, \Nc, $D$ (distance) and
$T$ (kinetic temperature) settles the others (Paper~I).
The column density is linked to the extinction profile also via the
\emph{gas-to-dust ratio} (we use this term as short form for the more precise
\emph{hydrogen column density per unit reddening by dust}).
Hence, if the BES model holds, 
one can determine the gas-to-dust ratio if $D$ and $T$ are
known, or one can determine the distance to the cloud by measuring $T$
and applying a ``standard'' gas-to-dust ratio. 
In any case, the
kinetic temperature remains the key parameter to measure.
A reliable temperature measurement in cold, dense cores is
possible using the 1.3~cm lines of
ammonia \citep{walmsley_83,danby_88}.
However, previous ammonia measurements 
\citep[\mbox{$T=16$~K},][]{bourke_95b}
are in contradiction to other temperature derivations
\citep[][Paper~I]{avery_87}. 

Here we present new ammonia measurements, carried out
with the Effelsberg 100-m telescope. Apart from the temperature
derivation and an assessment of the inherent uncertainties, we compare
the line-width to the width expected from purely thermal motion, which
is a crucial test of the assumption that B68 is in hydrostatic
equilibrium. We calculate the 
distance, gas-to-dust ratio and fractional ammonia abundance 
that follow from the BES
model and the new temperature value.

\section{Observations and data calibration\label{sec:observations}}

The \onetwo\ inversion lines of \NHt\ were observed simultaneously on
May, 6th, 2002 with the Effelsberg 100-m telecope (40\arcsec\ beam at
23.7~GHz). 
B68 was between 12.2\degree\ and 15.8\degree(max) elevation.
We used the new 1.3~cm HEMT (High Electron Mobility
Transistor) receiver in the frequency switching mode.
The system temperature including the atmosphere was 63--74~K. The
spectrometer was a 8192 channel autocorrelator split into four
quarters,
two for each polarization channel,
centred on the frequencies of the \onetwo\ transitions, 
\mbox{$\nu_0(1,1)=23.694495487$~GHz}
and \mbox{$\nu_0(2,2)=23.722633335$~GHz} respectively \citep[line
parameters from][]{kukolich67}. The velocity resolution was
0.062~\kmps.
Pointing was checked on NGC7027 and 
NRAO530 and was found to be better than 6\arcsec.

\begin{figure}
\includegraphics[width=\hsize,clip=false]{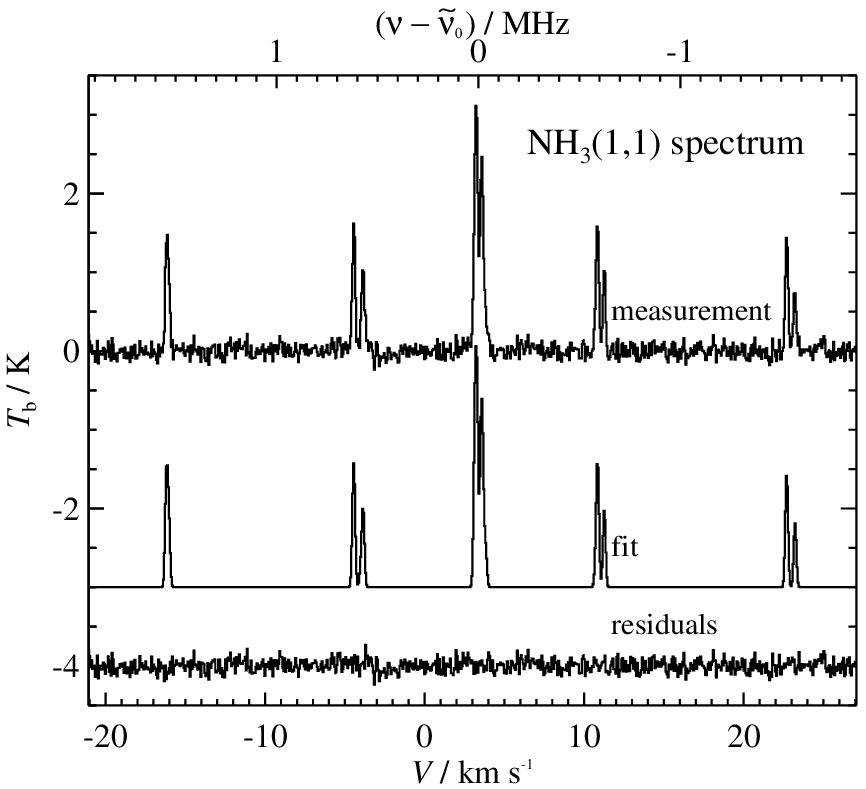}
\caption{B68 as observed in the \oneone\ inversion transition of \NHt.
The position observed is \eqeighttooo{17}{22}{46}{0}{-}{23}{49}{48}.
The y-axis gives the brightness temperature, \Tb. The upper x-axis gives the
frequency offset from the central frequency of the transition,
$\nu_0(1,1)$,
corrected for the relative velocity of the cloud. The velocity axis
corresponds to $\nu_0(1,1)$ and its zero point refers to 
$\nu_0(1,1)$ in the \mbox{$\Vlsr=0$} frame.
Top:
Calibrated spectrum after a total of 134 minutes effective integration
time.
Middle (shifted by $-3$~K for visibility): \chisq-fit assuming
LTE among the 18 hyperfine components and taking the velocity
resolution of the spectrometer into acccount.
Down (shifted by $-4$~K): Residual spectrum after
subtracting the fit.
\label{fig:oneone}}
\end{figure}

\begin{figure}
\includegraphics[width=\hsize,clip=false]{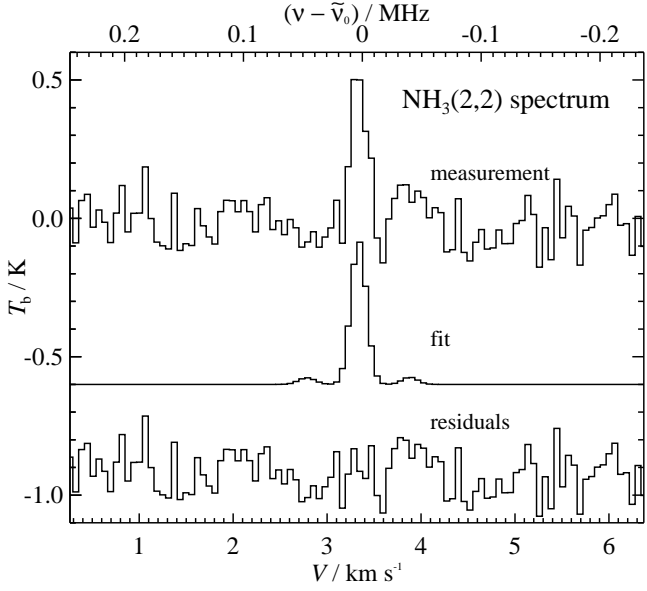}
\caption{B68 as observed in the \twotwo\ inversion transition of \NHt.
For observed position, integration time and description of plotting
axes see Fig.~\ref{fig:oneone}. 
Top:
Calibrated spectrum. The noise is still too high (and the temperature too low) 
to see the weaker hyperfine
components. 
Middle (shifted by $-0.6$~K for visibility): \chisq-fit, the fitted
paramters are $V_0$ and $\tau_0(2,2)$. The parameters
\Tex\ and \DV\ are kept fixed using the values derived in the $(1,1)$
fit.
Down (shifted by $-0.9$~K): Residual spectrum after
subtracting the fit.
\label{fig:twotwo}}
\end{figure}

The data were
calibrated using NGC7027, for which we assumed an observed flux
density of \mbox{$\Fnu(23.71~\mbox{GHz})=4.99$~Jy} \citep{ott_94}.
This value takes into 
account the 
spatial extension of the source
(true flux is 1.5\,\% higher)
and the epoch of the observation (annual flux decrease is 0.5\,\%).
We multiplied the spectra with a gain-elevation correction factor
based on the measurements of Altenhoff (1983, private communication),
which are consistent with more recent observations at intermediate
elevations provided by the telescope team.
The values for the
receiver sensitivity and main beam efficiency 
were 0.93~K/Jy and 58\,\% respectively.
The uncertainties of the calibration are (1) the uncertainty
of the absolute flux density of the calibration source,
(2) the statistical error of our calibration measurements
(changes in atmospheric conditions, telescope gain, etc.) and (3) the 
remaining uncertainty of
the gain-elevation dependency at very low elevations. They add up
to 20\,\%.
Summing, folding and baseline-fitting were done using CLASS.
\rv{Unfortunately, one of the polarization channels
showed differences between the on- and off-frequency spectra 
(possibly due to a sensitivity gradient of the receiver)
and also a variation of the spectra with time. 
As we could not reliably correct for these artifacts,
we decided to discard that channel in the following analysis.}
The calibrated spectra, obtained
with a total effective
integration time of 134 minutes, are shown in 
Figs.~\ref{fig:oneone} and~\ref{fig:twotwo}. The beam filling
factors assumed are \mbox{$\eta(1,1)=\eta(2,2)=1$}.

\section{Results\label{sec:results}}

The general concept of deriving cloud parameters from the hyperfine
spectra is presented
by \citet{ho_79}.
\NHt\ fundamentals can be found \eg\ in the papers of
\citet{kukolich67}, \citet{poynter_75}
and in the review by \citet{ho_83}.
In order to derive reliable values for the intrinsic line-width
we have convolved the model spectrum with the filter frequency
response function (approximated
by a gaussian with full width at half maximum (FWHM) of 
1~channel-width) 
when finding the best model spectrum in a \chisquare-fit to the
data.
Each channel was weighted with the inverse square of the rms
(root-mean-square) noise of the spectrum determined outside
the lines. For the $(1,1)$ spectrum, the four unknown variables
excitation temperature \Tex, total opacity $\tau_0(1,1)$, line-of-sight
velocity $V_0$ and line-width \DV\ can be determined from the
spectrum. The weaker hyperfine lines of the $(2,2)$ spectrum allow
only for the fitting of two unknowns, namely
$\tau_0(2,2)$ and $V_0$. We have assumed
\mbox{$\Tex(2,2)=\Tex(1,1)$} and \mbox{$\DV(2,2)=\DV(1,1)$}
\citep[for a discussion of these assumptions see][]{ho_79}, and
proceeded modelling the $(2,2)$ spectrum analogously to the $(1,1)$
case.
The modelled spectra are shown 
in Figs.~\ref{fig:oneone} and~\ref{fig:twotwo}, together with
the observed spectra and the residual spectra.
The parameters determined in the \chisquare-fits are given in
Table~\ref{tab:parameters}.

\begin{table}
\caption{Primary fitting parameters with formal errors
as determined in the
\chisquare-fits shown in Figs.~\ref{fig:oneone}
and~\ref{fig:twotwo}. We label $\tau_0$ as the \emph{total opacity} of
the transition; it is the the sum of the central opacities
of the individual hyperfine components
(all gaussian with \mbox{$\mbox{FWHM} = \DV$})
of the inversion transition in question.
The small discrepancy between the two transitions in the line-of-sight
velocities is also present in the (otherwise unused) second
polarization channel. Possible reasons are a mis-positioning of the
spectrometers (by 0.2 channels), an inaccuracy in the rest frequency
difference (of 2.8~kHz or 0.01\,\%) 
or a combination of rotation (of the globule) and
different opacities.
\label{tab:parameters}}
\begin{tabular*}{\hsize}{@{\extracolsep\fill}lll}
\hline 
                  & \multicolumn{2}{c}{Spectrum}\\
                  & \oneone     & \twotwo       \\
\hline 
\Tex\quad(K)      & $6.4\pm0.8$     & ---           \\
$\tau_0(J,K)$     & $7.2\pm0.4$     & $0.22\pm0.03$ \\
$V_0$\quad(\kmps) & $3.375\pm0.002$ & $3.34\pm0.02$ \\
\DV\quad(\kmps)   & $0.181\pm0.003$ & ---           \\
\hline 
\end{tabular*}
\end{table}

The rotational Temperature \Tot, characterising the ratio of the
populations in the \twotwo\ and \oneone\ rotational states, can be
calculated from
\begin{equation}
\Tot = \frac{-41.5~\K}{\ln\left(\frac{9}{20}\cdot\frac{\tau_0(2,2)}{\tau_0(1,1)}\right)}
\label{eq:trot}
\end{equation}
\citep[combining Eqs.~(2) and~(3) of][]{ho_79}. We find
\mbox{$\Tot=9.7\pm0.3$~K}. The partition function is calculated
assuming that only metastable \mbox{($J=K$)} rotational levels are
populated and \Tot\ is characteristic for all metastable levels and
the ground state. The \NHt\ column density calculated this way is 
\mbox{$\NNHt = \expm{5.4}{14}$~\pcms}.
According to \citet{walmsley_83} and
\citet{danby_88}, \Tot\ can be converted to the kinetic temperature,
$T$. At low temperature they are almost equal, and we get
\mbox{$T=9.9$~K}.
This temperature corresponds to a thermal line-width of
\mbox{$\DVtherm=0.164$~\kmps}
(\mbox{$\DVtherm\equiv\sqrt{8\ln\!2 (kT)/(17m_\mathrm{H})}$}, where
$k$ is the Boltzmann constant and $m_\mathrm{H}$ the atomic hydrogen mass).
See \citet{harju_93} for the equations used to calculate \NNHt\ and $T$.

The formal error of the temperature reflects only the rms noise.
The calibration error plays no role, because the two lines were
measured simultaneously and only their intensity \emph{ratio} enters
the temperature derivation. 
The true uncertainty of the kinetic
temperature must be assessed through a critical
review of the assumptions involved.
As discussed by \citet{ho_79}, the most serious errors are probably
the assumptions \mbox{$\Tex(1,1)=\Tex(2,2)$} and
\mbox{$\eta(1,1)=\eta(2,2)$}.
As peculiar excitation conditions are unlikely to
prevail in B68 (no internal heating
sources and no large temperature gradients expected), 
we expect these differences to be $\lesssim20\,\%$. 
Thus we have repeated our calculations
assuming $\pm20\,\%$ differences between the excitation temperatures
and between the beam filling factors of the two transitions. 
The temperature uncertainty introduced by these differences turns out
to be $+12/-9\,\%$. Including the formal error from the fit, we
finally derive 
\mbox{$T=9.9^{+1.3}_{-1.0}$~K}.

We also used the Monte Carlo radiative transfer program developed by
\citet{juvela97}
in order to test our previous approximation of
near-homogeneous excitation conditions along the line of sight
and this way checking our derived temperature value.
The level energies were calculated from the analytical fit of
\citet{poynter_75}, the radiative rates were calculated according to
formulae in \citet{townes_55} and the collisional coefficients were
taken from \citet{danby_88}.

As the absolute density of the underlying cloud model scales with 
the cloud's distance, we have run the program with several values
of $D$ (between 50 and 300~pc), each time optimising the
fractional abundance, the turbulent 
line-width and the kinetic temperature. The best fits were obtained
for \mbox{$T=9.7$~K} (independent of $D$).
The lowest \chisquare\ value was found for \mbox{$D=170$~pc}, but 
no accurate determination is possible as
values between 80 and 250~pc are all consistent with our
calibration uncertainty. 
We also tried non-isothermal cloud models (with unchanged density
structure)
by coupling the
kinetic temperature to the dust temperature \citep{zucconi_01}
in the inner region of the globule (defined by
\mbox{$n>n_\mathrm{crit}$}, trying
\mbox{$n_\mathrm{crit}=10^4,\;10^{4.5}\mbox{ and }10^5~\pcmc$}).
These temperature
distributions did \emph{not} improve the fits.

\section{Discussion\label{sec:discussion}}

\subsection{Consistency with other temperature values\label{sub:consistency}}

In Paper~I we concluded that most likely the kinetic temperature in
B68 is around 8~K. This was based on the measured \CeiO\onezero\ excitation
temperature of 8~K and our Monte Carlo modelling results.
Allowing for the possibility of subthermal excitation in the less
dense outer parts, where most of the CO emission comes from, the
kinetic temperature derived here is in agreement with our earlier CO
measurements. 
\citet{avery_87} deduced from their CO and \thCO\
observations an outward increasing kinetic temperature between 6 and
11~K. As they used a constant fractional CO abundance, the derived
gradient must be regarded with caution. However,
the given range there covers the temperature derived here.
Temperatures derived from ammonia in other globules without internal
heating sources often lie around 10~K \citep{lemme_96}.
\rv{\citet{galli_02} calculated the gas temperature distributions in 
molecular cloud cores and predicted for B68 a temperature of about
10~K, increasing only slightly towards the cloud edge.}

There are two other attempts to derive the kinetic temperature of B68
by means of dedicated \NHt\ inversion line observations: \citet{bourke_95b}
derived \mbox{$T=16$~K}.
using the same underlying assumptions as we have used. As discussed in
Paper~I, this derivation is likely to involve some unfortunate error
either in the calibration process or in the calculation.
Very recently, \citet{lai_02} have presented an estimate of the ammonia
rotational temperature, which
corresponds to a kinetic temperature
of \mbox{$T=11.2\pm0.9$~K}, 
which is consistent with our value.

\subsection{Turbulence\label{sub:turbulence}}

The thermal line-width \mbox{$\DVtherm=0.164\pm0.010~\kmps$}
is only marginally smaller than the actual line-width
\mbox{$\DV=0.181\pm0.003~\kmps$}, which is an 
essential condition for the
BES premise of hydrostatic equilibrium.
The internal support provided by turbulence is
\begin{equation}
E_\mathrm{turb} = \frac{2.33}{17}\left(\frac{\DV^2}{\DVtherm^2} -
1\right)E_\mathrm{therm}\;,
\label{eq:eturb}
\end{equation}
where $E_\mathrm{therm}$ is the thermal energy and 
$2.33m_\mathrm{H}$ is the mean molecular weight we assume throughout
this paper. 
From our measurements we derive
\mbox{$E_\mathrm{turb} / E_\mathrm{therm} = 3\,\%$}.
The small contribution of the macroscopic motion is well below the
uncertainty of the thermal energy and therefore it can be neglected
as a physical parameter of the BES model.

\subsection{Fixing the BES model\label{sub:fixing}}

As mentioned in Sect.~\ref{sec:introduction}, after
determination of the kinetic temperature the BES model of B68 is fixed if
only one of the parameters \nc, $D$ or gas-to-dust ratio is known.
Using Eqs.~(2) and~(11) of Paper~I we get for \mbox{$T=10$~K}
\begin{equation}
\begin{array}{rcl}
\Nc     & = & \expm{1.79}{22}~\pcms   \,   \left(\frac{\nc}{10^5\,\pcmc}\right)^{1/2}\\
\Nc     & = & \expm{2.59}{22}~\pcms   \,   \left(\frac{D}{100\,\pc}\right)^{-1}\\
        & \mbox{and} &                        \\
\Nc     & = & \expm{3.05}{22}~\pcms    \,   \left(\frac{N(\mathrm{H_2})/\EHK}{\expm{1.23}{22}\,\pcms\,\pmag}\right)\;.\\
\end{array}
\label{eq:Ncs}
\end{equation}
In the third equation (the only one which does not require the BES
assumptions to hold) we have assumed \mbox{$N/N(\Htwo)=6/5$}, and the
term in brackets corresponds to the standard gas-to-dust ratio
\citep{bohlin_78,cardelli_89}.

The number density of molecular hydrogen can be estimated from \NHt\
data by balancing collisional excitation against emission
\citep{ho_83}. However, to derive a value close to the \emph{central}
density, a high spatial resolution is required, which cannot
be achieved with single-dish observations.
As B68 has no detectable foreground stars, its distance is only known
as far as its association with the 
Ophiuchus complex holds. According to \citet{degeus_89} this molecular
cloud complex extends from \mbox{$D=80\mbox{ to }170$~pc}
with a central value of 125~pc.
The gas-to-dust ratio is expected to vary from
cloud to cloud, but
there are a
number of measurements suggesting that the variation does not exceed a
factor of 2 (Paper~I).

The relation between the distance and
the gas-to-dust ratio imposed on B68 by the BES model reads
\begin{equation}
D = 85~\pc\,\left(\frac{N(\mathrm{H_2})/\EHK}{\expm{1.23}{22}\,\pcms\,\pmag}\right)^{-1}\;.
\label{eq:distance}
\end{equation}
The globule is thus located on the near side of the Ophiuchus
complex unless B68 has a \emph{smaller} than average gas-to-dust
ratio.
The latter
would be unexpected given
the common understanding of dust evolution in cold and
dense environments \citep{kim_96}. However, our Monte Carlo
simulations preferred a distance in the 100--200~pc range
so this possibility should not be ruled out.
For the standard gas-to-dust ratio
the mass
and the external pressure of the BES are 
\mbox{$M =0.9~\Msun\,\,(D/85\,\pc)$}
and
\mbox{$\PR =\expm{2.3}{-12}~\mathrm{Pa}\,\,(D/85\,\pc)^{-2}$}
respectively.

\subsection{Ammonia abundance\label{sub:abundance}}

The beam averaged ammonia column density has been calculated in
Sect.~\ref{sec:results}. In order to get the fractional abundance
\mbox{$\chin=N(\NHt)/N(\Htwo)$}, we have convolved the BES column density
profile with a 40\arcsec\ (FWHM) gaussian. Taking into account that
the observed position is 19\arcsec\ off 
the position of maximum extinction 
\citep[as reported by][]{bergin_02}, the column density towards the
peak extinction position is \mbox{$N(\NHt)=\expm{7.7}{14}~\pcms$}, and
with Eqs.~\ref{eq:Ncs} and~\ref{eq:distance} we get 
\mbox{$\chin=\expm{3.0}{-8}\,\,(D/85\,\pc)$}.
This value is almost equal to the median value of a sample of 22
ammonia clumps in Orion \citep{harju_93} and very close to the
fractional abundance in \object{B217SW}, a Taurus dense core with
similar temperature and size as B68 \citep{hotzel_01}.

\section{Conclusions\label{sec:conclusions}}

Our observations support the hypothesis that B68 is in a state of
isothermal hydrostatic equilibrium. Its kinetic temperature is
\mbox{$10\pm1.2$~K} and its turbulent support is negligible.
The ammonia abundance is close to the values found in other dark
cores, but from the BES scaling relations we conclude that either the
distance or the gas-to-dust ratio of B68 is smaller than previously
expected.

\begin{acknowledgements}

We thank Dr.~Floris van der Tak for providing the \NHt\ collisional
coefficients in electronic form.
This project was supported by
the Academy of Finland, grant
nos.~173727 and 174854.

\end{acknowledgements}

\end{document}